\documentstyle[preprint,aps]{revtex}
\begin{document}
\draft
\title
{Assumption of nonvanishingness of vacuum expectation of the scalar
field for spontaneous symmetry breaking is superfluous
}
\author{Yu Shi\cite{email}}
\address{Department of Physics, Bar-Ilan University, Ramat-Gan 52900, Israel}
\maketitle
\begin{abstract}
For spontaneous breaking of global or gauge symmetry, it is superfluous
to assume that the  vacuum expectation value of the scalar field  
manifesting the symmetry is nonvanishing. The vacuum with spontaneous
symmetry breaking simply
corresponds to the nonzero number of particles 
of one or more components of the real scalar field.
\end{abstract}
\widetext
A crucial ingredient of the theories of spontaneous symmetry breaking
is  the assumption that
 the  vaccum expectation  value of the scalar field manifesting the the 
 symmetry, $\overline{\phi}$, is nonzero 
 \cite{weinberg,goldstone,gsw,higgs,hooft,ws}.
  This conflicts  the
 basic fact that the Hilbert space is spanned by states with definite number
  of particles and/or antiparticles in each mode. 
  In fact, $\overline{\phi}\,\neq\,0$ is only a sufficient
  but not a necessary condition for the required results. Though it  is widely
 acknowledged  as a property of vacuum, here we point out that this assumption 
  is superfluous. 
  For spontaneous gauge symmetry breaking in 
  superfluidity or superconductivity, it  has been pointed out that
  the nonvanishingness of the expectation value of the field operator is
  by no means essential but an approximate though convenient approach
  \cite{shi}.
 
 We follow the  notations and line of development 
 in \cite{weinberg}, and start
 with Goldstone theorem. Under a continuous symmetry which transforms 
a set of Hermitian scalar fields $\phi_{n}(x)$ as
\begin{equation}
\phi_{n}(x)\,\rightarrow\, \phi_{n}(x)+i\epsilon\sum_{m}t_{nm}\phi_{m}(x),
\label{tran}
\end{equation}
where $it_{nm}$ is the finite and real 
matrix corresponding to the symmetry transformation. Consequently, 
the invariance of the action and measure, and thus the effective potential
leads to
\begin{equation}
\sum_{nm}\frac{\partial V(\phi)}{\partial \phi_{n}}t_{nm}\phi_{m}
\,=\,0,
\end{equation}
therefore at the minimum the $V(\phi)$,
\begin{equation}
\sum_{nm}\Delta^{-1}_{nl}(0)t_{nm}\phi_{m}\,=\,0,	\label{gold}
\end{equation}
where 
$\Delta^{-1}_{nl}(0)\,=\,\partial^{2}V(\phi)/\partial\phi_{n}\partial\phi_{l}$
is the reciprocal of the momentum space propagator. In the 
conventional
 approach, the indication of symmetry breaking
 is $\sum_{m} t_{nm}\overline{\phi_{m}}\,\neq\,0$ obtained from
(\ref{tran}).  By calcualting the vacuum expectation
of (\ref{gold}), one obtains Goldstone theorem,
i.e. $\sum_{m} t_{nm}\overline{\phi_{m}}\,\neq\,0$ implies the
massless boson. There is one massless boson for every independent 
broken symmetry.

 Now we insist that
  $\overline{\phi_{m}}$, and thus the expectation of each term in 
  ({\ref{tran}) vanishes, therefore    the expectation of
 (\ref{gold}) is trivially satisfied.
 To examin the consequence of symmetry breaking, one should
 study another operator which is a functional of $\phi_{n}$ and can also
 reflect the symmetry of the Hamiltonian, or say the 
 effective potential.
 This is just $\phi_{n}\phi_{n}$.
For instance, in the classical example
 \begin{equation}
 {\cal L} \,=\,-\frac{1}{2}\sum_{n}\partial_{\mu}\phi_{n}
 \partial^{\mu}\phi_{n}
  -\frac{{\cal M}^{2}}{2}\sum_{n}\phi_{n}\phi_{n}
  -\frac{g}{4}(\sum_{n}\phi_{n}\phi_{n})^{2}, \label{on}
  \end{equation}
  which is invariant under the group $O(N)$, the 
  effective potential is a functional of $\phi_{n}\phi_{n}$.

Under the transformation (\ref{tran}),
$\phi_{n}\phi_{n}$ transforms as
\begin{equation}
\phi_{n}(x)\phi_{n}(x)\,\rightarrow\, 
\phi_{n}\phi_{n}(x)+2i\epsilon\sum_{m}t_{nm}\phi_{n}(x)\phi_{m}(x).
\label{tran2}
\end{equation}
If the symmetry is not broken,  $\overline{\phi_{n}\phi_{n}}$ 
remains unchanged.  Since $\overline{\phi_{n}\phi_{m}}\,\equiv\,0$
for $n\,\neq\,m$, (\ref{tran2}) implies 
$\overline{\phi_{n}\phi_{n}}\,\rightarrow\,(1+2\epsilon i)\overline{\phi_{n}\phi_{n}}$,
therefore $\overline{\phi_{n}\phi_{n}}$ and thus the 
effective potential is invariant if and only if
$\overline{\phi_{n}\phi_{n}}\,=\,0$.

On the other hand, the mode expansion
\begin{equation}
\phi_{n}\,=\,\int \frac{d^{3}k}{(2\pi)^{3}}\frac{1}{\sqrt{2\omega_{k}}}
(a_{k}^{(n)}e^{ikx}+{a_{k}^{(n)}}^{\dagger}e^{-ikx}) \label{expan}
\end{equation}
with $[a_{k}^{(n)},{a_{k'}^{(n)}}^{\dagger}]\,=\,(2\pi)^{3}\delta^{(3)}(k-k')$ 
yields
\begin{equation}
\overline{\phi_{n}\phi_{n}}\,=\,
\int \frac{d^{3}k}{(2\pi)^{3}}\frac{1}{\omega_{k}} \overline{N_{k}^{(n)}}+C,
\label{phi}
\end{equation}
where $N_{k}^{(n)}\,=\,{a_{k}^{(n)}}^{\dagger}a_{k}^{(n)}$,
$C\,=\,\int d^{3}k/2\omega_{k}$ is an infinite c-number.
In the derivation, we exploited the property that 
$\overline{a_{k}^{(n)}a_{k}^{(n)}}\,=\,0$ and 
$\overline{a_{k}^{(n)}{a_{k'}^{(n)}}^{\dagger}}\,=\,0$ for $k\,\neq\,k'$.
Note that the value of
 $\overline{\phi_{n}\phi_{n}}$ is independent of $x$.

 The least value of  (\ref{phi}) correponds to $N_{k}^{(n)}\,=\,0$.
 To reconcile with 
 $\overline{\phi_{n}\phi_{n}}\,=\,0$ in the absence of symmetry breaking,
 the term of 
 infinite c-number should be ignored. As well known,
 the same problem and strategy appear in the 
 energy calculated from (\ref{expan}). In conclusion, the
 vacuum without symmetry breaking
 is simply the state in which there is no particle  and 
 thus $\overline{\phi_{n}\phi_{n}}\,=\,0$.
 
 Eq. (\ref{tran2}) implies that 
 spontaneous symmetry breaking corresponds to
 $\sum_{m}t_{nm}\overline{\phi_{n}\phi_{m}}\,\neq\,0$,
 therefore $\overline{\phi_{n}\phi_{n}}\,\neq\,0$, 
 it is random for which $n$ 
 this enequality holds, as implied by symmetry.
 
 Multiplying (\ref{gold}) by $\sum_{n}\phi_{n}$ and considering 
 $\overline{\phi_{n}\phi_{m}}\,\equiv\,0$ for $n\,\neq\,m$,
 we obtain
 \begin{equation}
\sum_{nm}\Delta^{-1}_{nl}(0) t_{nm}\overline{\phi_{m}\phi_{m}}\,=\,0.
\label{gold2}
\end{equation}
Therefore whenever there is a component $k$ for which
$\overline{\phi_{k}\phi_{k}}$ is nonzero, 
the summation $\sum_{m}t_{nm}\overline{\phi_{m}\phi_{m}}$ 
which includes $\overline{\phi_{k}\phi_{k}}$ must be nonzero sice
each $\overline{\phi_{m}\phi_{m}}\,\leq\,0$. The nonvanishing
 $\sum_{m}t_{nm}\overline{\phi_{m}\phi_{m}}$ should be an eigenvector
 of $\Delta^{-1}_{nl}(0)$ with
 eigenvalue zero, thus  $\Delta_{nl}(q)$ has a pole at $q^{2}\,=\,0$, i.e. there
  is a massless boson.
  There is one massless boson  for every independent broken symmetry.
 All essential results of Goldstone theorem can thus obtained.

For  the  example (\ref{on}), the
 minimum of the effective potential is at
\begin{equation}
\sum_{n}\overline{\phi_{n}\phi_{n}}\,=\,-\frac{{\cal M}^{2}}{g},
\end{equation}
the mass matrix is
\begin{eqnarray}
M^{2}_{nm} & = & \overline{\frac{\partial^{2}V(\phi)}
{\partial \phi_{n}
\partial \phi_{m}}}\nonumber \\
 & = & 2g\overline{\phi_{n}\phi_{n}}\delta_{nm}.
 \end{eqnarray}

 $O(N)$ symmetry is broken down to $O(N-1)$  when there
is one component $\phi_{1}$ with $\overline{\phi_{1}\phi_{1}}\,\neq\,0$
while $\overline{\phi_{i}\phi_{i}}\,=\,0$ for $i\,\neq\,1$, therefore
$\overline{\phi_{1}\phi_{1}}\,=\,
-{\cal M}^{2}/g$.
Consequently there is one massive boson
with mass $2|{\cal M}^{2}|$ and $N-1$ massless bosons. 
 This is consistent with the general argument above. The massless bosons 
correpond to the $O(N-1)$ symmetry relating different possible selection
of $\phi_{1}$.

From the general viewpoint of quantum mechanics, if initially the state of a
system is in an eigenstate of a relevant operator or
 a set of operators
commuting each other, it will  always be in this stationary state if these
operators commute the Hamiltonian, or in a nearly-stationary
state if there is near-degeneracy in case the relevant 
operator does not commute the Hamiltonian. 
The initial condition is determined by a basic postulate that
the measurement projects the state to an eigenstate of the 
relevant operator {\em defining} the physical situation. 
We think this is the essence of 
various spontaneous symmetry breaking \cite{shi}. 
In the present case,
the relevant operators are $\phi_{n}\phi_{n}$, which commute 
the effective
 potential. Therefore spontaneous summetry breaking can occur.
 Indeed,  $\phi_{n}\phi_{n}$ represents the particle numbers in all modes,
 and the vacuum or excited state is just defined through the
 particle number.
  Of course,
as well known,
 the results also applies if the symmetry is broken non-spontaneously.
The vacuum is nothing but  the ground state, with least total
number of particles,
spontaneous
symmetry breaking makes the state of the system 
in one of the degenerate ground states 
 instead of the combination, therefore
the number of
 particles of one or more components of the scalar field are nonzero,
In the meantime, massless bosons are yielded with number determined by the
broken symmetry.

In the  gauge symmetry breaking,  the gauge field 
acquires a mass, this can also be obtained from the present argument.
Defining $v_{n}\,=\,\sqrt{\overline{\phi_{n}(0)\phi_{n}(0)}}$, the new
fields $\tilde{\phi}(x)$ in the unitary gauge
can be obtained by  $\sum_{nm}\tilde{\phi}(x)t_{nm}v_{m}\,=\,0$.
Substituting the  shift field 
$\phi'_{n}\,=\,\phi_{n}-v_{n}$ 
to the
Lagrangian
\begin{equation}
{\cal L}\,=\,-\frac{1}{2}\sum_{n}(\partial_{\mu}\tilde{\phi}_{n}-i\sum_{m\alpha}
t^{\alpha}_{nm}A_{\alpha\mu}\tilde{\phi}_{m})^{2},
\end{equation}
the gauge boson masses $\mu^{2}_{\alpha\beta}\,=\,-\sum_{nml}t^{\alpha}_{nm}
t^{\beta}_{nl}v_{m}v_{l}$ is yielded in the same way as the 
previous approach.
The proof of renormalizability is also
 valid  with the new definition of
$v_{n}$ here. In the $SU(2)\times U(1)$ electroweak theory,
the doublet scalar field
can be in the form of
\begin{equation}
\phi=\left( \begin{array}{c}
  0\\ \phi_{0}
  \end{array} \right)
\end{equation}
with $\phi_{0}$ hermitian. Defining the shift fields as
\begin{equation}
\phi=\left(\begin{array}{c}
  0\\ \sqrt{\overline{\phi_{0}\phi_{0}}}
  \end{array}\right)+ 
\left(\begin{array}{c}
  \phi'_{+}\\ \phi'_{0}
    \end{array}\right),
 \end{equation}
all the previous results are yielded in the similar way.

E. J. Weinberg is acknowledged for comment on an earlier version of this paper.

 \end{document}